# Enhanced entanglement of two optical modes in optomechanical systems via an optical parametric amplifier

**Rong-Guo Yang**[1,2,4*], **Ni Li**[1,2], **Jing Zhang**[1,2,4], **Jie Li**[1,3,4,5], **Tian-Cai Zhang**[1,3,4]

[1] State Key Laboratory of Quantum Optics and Quantum Optics Devices, Shanxi University, Taiyuan 030006, China

[2] College of Physics and Electronic Engineering, Shanxi University, Taiyuan 030006, China

[3] Institute of Opto-Electronics, Shanxi University, Taiyuan 030006, China

[4] Collaborative Innovation Center of Extreme Optics, Shanxi University, Taiyuan 030006, China

[5] School of Science and Technology, Physics Division, University of Camerino, I-62032 Camerino (MC), Italy

E-mail: yrg@sxu.edu.cn

## Abstract

We investigate the effect of a degenerate optical parametric amplifier (OPA) placed inside an optomechanical cavity on the steady-state entanglement of two cavity modes, which jointly interact with a mechanical resonator. Two cavity modes are respectively driven at the red and blue sideband associated with the mechanical resonator, which generates entanglement between them in the limit of resolved sideband. The OPA gives rise to single-mode squeezing of the cavity fields, which results in significant improvement of the two-mode entanglement. It is found that an optimal nonlinear gain of the OPA exists, depending on the system temperatures, which yields the maximum entanglement. The improvement is particularly remarkable for the system at cryogenic temperatures.

## 1. Introduction

Quantum entanglement is a valuable resource that can be exploited not only in quantum information processing, e.g., performing computation and communication tasks with an efficiency which is not achievable classically [1], but also in the study of the quantum-to-classical transition [2, 3], wave-function collapse theories [4-6], and so on. To date, great efforts have been made on the demonstration of entanglement in microscopic systems. However, entanglement in macroscopic systems has been less investigated and observed.

Optomechanics, exploring the interaction between light and mechanical objects via radiation pressure [7], is considered as an ideal platform to prepare entangled states, especially of large and massive objects [8]. In the past two decades, many efforts have been made to prepare entangled states in cavity optomechanical systems. Basically, they can be divided into the following kinds: entangled states of cavity modes [9-14], of a cavity mode and a mechanical mode [15-17], of mechanical modes [18-26], and of hybrid modes, e.g., in atom-optomechanical systems [27-30]. Recently, it has been shown that, by including an optical parametric amplifier (OPA) inside the cavity, interesting phenomena would occur. The OPA is able to enhance optomechanical cooling [31], optomechanical coupling strength and normal mode splitting [32]. The enhanced coupling strength makes it even possible to implement cavity optomechanics in the single-photon strong coupling regime [33]. The OPA generates squeezing of the cavity field which can be used to improve the sensitivity of mechanical quadrature measurements [34], and to prepare squeezed states of the mechanical mode [35]. Moreover, it has been shown that the OPA can also enhance the entanglement of one cavity mode and one mechanical mode [36], of two mechanical modes [37], and of multi cavity and mechanical modes [38].

In the present paper, we provide a protocol to enhance the stationary continuous variable entanglement between two cavity modes via placing a degenerate OPA inside an optomechanical cavity, which is comprised of a fixed mirror and a light movable mirror which acts as a mechanical resonator. The two cavity modes jointly interact with the mechanical resonator and by properly choosing the cavity-laser detunings the two cavity modes can be prepared in an entangled state. The OPA is used to squeeze both the two cavity fields, and as a result, the entanglement between the two cavity modes can be significantly enhanced. We focus on the case of entanglement in steady states. It is found that an optimal nonlinear gain of the OPA and an optimal phase of the optical field driving the OPA exist corresponding to the maximum entanglement and this optimal nonlinear gain becomes smaller as the system temperature increases. This implies it is not true that larger the single-mode squeezing corresponds to stronger the two-mode entanglement. One has to optimize the nonlinear gain of the OPA so as to get the maximum entanglement at specific temperatures.

The paper is organized as follows: In Section 2 we introduce in detail our model and provide the system Hamiltonian and its corresponding quantum Langevin equations (QLEs) after linearization of the system dynamics. In Section 3, we present the results and compare the entanglement of two cavity modes in the cases with and without inserting an OPA. It shows that remarkable improvement of the steady-state entanglement can be achieved with the presence of the gain medium. Finally, we draw our conclusions in Section 4.

## 2. System Hamiltonian and Langevin equations

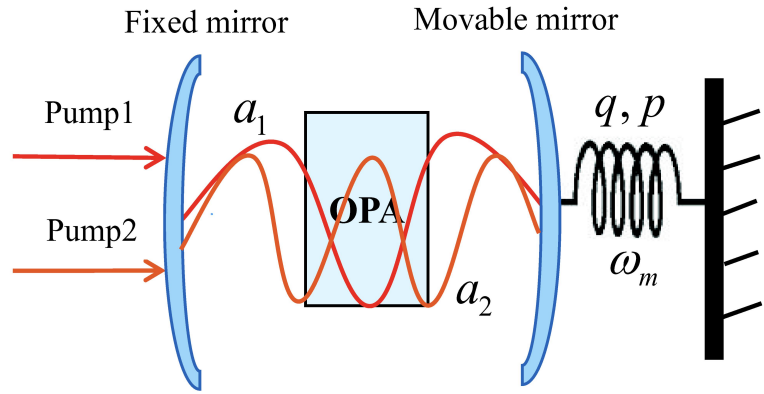

**Figure 1.** Sketch of the system. Two cavity modes are respectively driven by two lasers which simultaneously interact with the movable mirror via optomechanical interaction. A nonlinear crystal (OPA) inside the cavity is used to squeeze the cavity fields by optical parametric process.

As depicted in Fig. 1, we consider an optical Fabry-Perot cavity within which a degenerate OPA is placed. One cavity mirror is fixed and the other one is movable, which is modeled as a quantum mechanical harmonic oscillator with effective mass $m$ and frequency $\omega_m$. We consider two optical modes of the cavity with resonance frequencies $\omega_{Cj}$ ($j$=1,2), which are respectively driven by two lasers with frequencies $\omega_{Lj}$. The degenerate OPA is pumped by another two lasers at frequencies $2\omega_{Lj}$, which is used to generates two squeezed optical fields at frequencies $\omega_{Lj}$. We assume that the two frequencies $\omega_{L1}$ and $\omega_{L2}$ are very close

$|\omega_{L1} - \omega_{L2}| \approx n\Delta_C$ ($n$ is an integer, e.g., $n=10$), where $\Delta_C = \frac{\pi c}{L}$ ($c$ is the speed of light and $L$ is the cavity length) is the free spectral range of the cavity. For $L \sim 1\,\text{cm}$, $\Delta_C$ is about $\sim 10^{11}$ Hz, which is much smaller than the optical frequencies. In this case, the two pump fields interact synchronously and independently with the OPA and the two nonlinear processes can be considered identical. The two cavity modes interact via the usual optomechanical interaction with the mechanical resonator. The Hamiltonian for such a system is given by

$$\hat{H} = \sum_{i=1}^{2} \hbar\omega_{Ci} a_i^\dagger a_i + \frac{\hbar\omega_m}{2}(q^2 + p^2) - \sum_{i=1}^{2} \hbar g_i a_i^\dagger a_i q + \sum_{i=1}^{2} i\hbar\varepsilon_i \left(a_i^\dagger e^{-i\omega_{Li}t} - a_i e^{i\omega_{Li}t}\right) \\ + \sum_{i=1}^{2} i\hbar G \left(e^{i\theta} a_i^{\dagger 2} e^{-i2\omega_{Li}t} - e^{-i\theta} a_i^2 e^{i2\omega_{Li}t}\right), \quad (1)$$

where $a_i$ and $a_i^\dagger$ ($[a_i, a_j^\dagger] = \delta_{ij}$, $i,j=1,2$) are, respectively, the annihilation and creation operators of the cavity mode with frequency $\omega_{Ci}$. $q$ and $p$ ($[q,p]=i$) are the dimensionless position and momentum operators of the mechanical resonator. $g_i = (\omega_{Ci}/L)\sqrt{\hbar/m\omega_m}$ is the single-photon optomechanical coupling associated with the cavity mode with frequency $\omega_{Ci}$. $\varepsilon_i$ is the coupling between the driving laser and the cavity field, which is related to the pump power $P_i$ and the cavity decay rate $\kappa$ by $\varepsilon_i = \sqrt{2\kappa P_i/\hbar\omega_{Li}}$, where $\kappa = \frac{\pi c}{2FL}$ with $F$ the cavity finesse. The last term in the Hamiltonian is the novel part, which denotes the coupling between the OPA and the two cavity modes. $G$ is the nonlinear gain of the OPA, which is proportional to the power of the driving field, and $\theta$ is the phase of the optical field driving the OPA. Without loss of generality, we have assumed the nonlinear gain $G$ and the phase $\theta$ are identical for the two independent nonlinear processes.

The corresponding nonlinear QLEs, including various noises entering into the system, in the interaction picture with respect to $\hbar\omega_{Li} a_i^\dagger a_i$, are given by

$$\dot{q} = \omega_m p,$$
$$\dot{p} = -\omega_m q - \gamma_m p + g_1 a_1^\dagger a_1 + g_2 a_2^\dagger a_2 + \xi, \quad (2)$$
$$\dot{a}_1 = -(\kappa + i\Delta_{01})a_1 + ig_1 a_1 q + \varepsilon_1 + 2Ge^{i\theta} a_1^\dagger + \sqrt{2\kappa}\, a_1^{in},$$

$$\dot{a}_2 = -(\kappa + i\Delta_{02})a_2 + ig_2 a_2 q + \varepsilon_2 + 2Ge^{i\theta}a_2^\dagger + \sqrt{2\kappa}a_2^{in},$$

where $\Delta_{0i} = \omega_{Ci} - \omega_{Li}$, $\gamma_m$ is the mechanical damping rate, $\xi$ is the Langevin force operator accounting for the Brownian motion of the mirror, which is auto-correlated as [39]

$$\langle \xi(t)\xi(t')\rangle = \frac{\gamma_m}{\omega_m}\int \frac{d\omega}{2\pi}e^{-i\omega(t-t')}\omega\left[\coth\left(\frac{\hbar\omega}{2k_B T}\right)+1\right], \quad (3)$$

where $k_B$ is the Boltzmann constant and $T$ is the environmental temperature. $a_i^{in}$ is the input vacuum noise operator for the cavity, of which the only nonzero correlation function is

$$\langle a_i^{in}(t)a_i^{in\dagger}(t')\rangle = \delta(t-t'). \quad (4)$$

Sizeable steady-state entanglement is typically achieved with sufficiently large optomechanical couplings, which is realized when the cavity is intensely driven so that the intracavity field is strong. In this case, it is appropriate to focus on the linearized dynamics of the quantum fluctuations around the classical average values. For this purpose, one can write $a = \alpha_s + \delta a$, $q = q_s + \delta q$ and $p = p_s + \delta p$, and insert them into the QLEs of Eq. (2). The corresponding average values are obtained by setting the derivatives to zero, which are

$$p_s = 0, \quad q_s = \frac{g_1|\alpha_{s1}|^2 + g_2|\alpha_{s2}|^2}{\omega_m},$$

$$\alpha_{s1} = \frac{\varepsilon_1}{(\kappa - 2G\cos\theta) + i(\Delta_1 - 2G\sin\theta)}, \quad \alpha_{s2} = \frac{\varepsilon_2}{(\kappa - 2G\cos\theta) + i(\Delta_2 - 2G\sin\theta)}, \quad (5)$$

where $\Delta_1 = \Delta_{01} - g_1 q_s$, $\Delta_2 = \Delta_{02} - g_2 q_s$ are the effective cavity detunings including the frequency shift due to the interaction with the mechanical resonator. We see that the presence of the OPA leads to two effects: it modifies the cavity decay rate $\kappa \to \kappa - 2G\cos\theta$ and also the effective detunings $\Delta_{1,2} \to \Delta_{1,2} - 2G\sin\theta$. The corresponding QLEs for the quantum fluctuations of the system, after linearization of the dynamics around those steady state values, are given by

$$\delta\dot{q} = \omega_m \delta p,$$
$$\delta\dot{p} = -\omega_m \delta q - \gamma_m \delta p + G_1 \delta X_1 + G_2 \delta X_2 + \xi,$$
$$\delta\dot{X}_1 = -(\kappa - 2G\cos\theta)\delta X_1 + (\Delta_1 + 2G\sin\theta)\delta Y_1 + \sqrt{2\kappa}X_1^{in},$$

$$\delta \dot{Y}_1 = -(\kappa + 2G\cos\theta)\delta Y_1 - (\Delta_1 - 2G\sin\theta)\delta X_1 + G_1\delta q + \sqrt{2\kappa}Y_1^{in}, \quad (6)$$

$$\delta \dot{X}_2 = -(\kappa - 2G\cos\theta)\delta X_2 + (\Delta_2 + 2G\sin\theta)\delta Y_2 + \sqrt{2\kappa}X_2^{in},$$

$$\delta \dot{Y}_2 = -(\kappa + 2G\cos\theta)\delta Y_2 - (\Delta_2 - 2G\sin\theta)\delta X_2 + G_2\delta q + \sqrt{2\kappa}Y_2^{in},$$

where we have defined the quadrature fluctuation operators of the cavity modes $\delta X_i = (\delta a_i + \delta a_i^\dagger)/\sqrt{2}$, $\delta Y_i = i(\delta a_i^\dagger - \delta a_i)/\sqrt{2}$ and the corresponding input noise operators $X_i^{in} = (a_i^{in} + a_i^{in\dagger})/\sqrt{2}$, $Y_i^{in} = i(a_i^{in\dagger} - a_i^{in})/\sqrt{2}$. $G_i = \sqrt{2}g_i\alpha_{si}$ is the effective optomechanical coupling strength, where we have taken $\alpha_{si}$ real by properly choosing the phase reference of the cavity fields.

The above QLEs (6) can be rewritten in the following form

$$\dot{u}(t) = Au(t) + n(t), \quad (7)$$

where $u(t) = (\delta q, \delta p, \delta X_1, \delta Y_1, \delta X_2, \delta Y_2)^T$ is the vector of quadrature fluctuation operators, $A$ is the so-called drift matrix, which takes the form of

$$A = \begin{pmatrix} 0 & \omega_m & 0 & 0 & 0 & 0 \\ -\omega_m & -\gamma_m & G_1 & 0 & G_2 & 0 \\ 0 & 0 & -\kappa + 2G\cos\theta & \Delta_1 + 2G\sin\theta & 0 & 0 \\ G_1 & 0 & -\Delta_1 + 2G\sin\theta & -\kappa - 2G\cos\theta & 0 & 0 \\ 0 & 0 & 0 & 0 & -\kappa + 2G\cos\theta & \Delta_2 + 2G\sin\theta \\ G_2 & 0 & 0 & 0 & -\Delta_2 + 2G\sin\theta & -\kappa - 2G\cos\theta \end{pmatrix}, \quad (8)$$

and $n(t) = (0, \xi(t), \sqrt{2\kappa}X_1^{in}(t), \sqrt{2\kappa}Y_1^{in}(t), \sqrt{2\kappa}X_2^{in}(t), \sqrt{2\kappa}Y_2^{in}(t))^T$ is the vector of noise quadrature operators associated with the noise terms in Eq. (6). The system is stable when all the eigenvalues of the drift matrix $A$ have negative real parts. Since we are interested in the entanglement of two optical modes in steady state, all the results throughout the paper are presented with this stability condition fulfilled.

Due to the Gaussian nature of the quantum noise terms in Eq. (7) and the linearized dynamics, the steady-state quantum fluctuations of the system is a tripartite Gaussian state of two optical modes and one mechanical mode, fully characterized by the 6×6 covariance matrix

$V$ with its entries defined as $V_{ij}=(\langle u_i(\infty)u_j(\infty)+u_j(\infty)u_i(\infty)\rangle)/2$. The steady state covariance matrix $V$ can be obtained by solving the Lyapunov equation

$$AV+VA^T=-D, \qquad (9)$$

where $D$ is the diffusion matrix, with its entries defined as

$$\frac{1}{2}\langle n_i(t)n_j(s)+n_j(s)n_i(t)\rangle = D_{ij}\delta(t-s). \qquad (10)$$

The diffusion matrix is a diagonal matrix, that is $D=\mathrm{diag}\left[0,\gamma_m(2\bar{n}+1),\kappa,\kappa,\kappa,\kappa\right]$. Note that we have assumed the mechanical resonator is of high quality factor $Q=\omega_m/\gamma_m \gg 1$, which is typically satisfied under the current experiment conditions [40]. In this limit, $\xi(t)$ becomes $\delta$-correlated, i.e.

$$\langle \xi(t)\xi(t')+\xi(t')\xi(t)\rangle/2 \simeq \gamma_m(2\bar{n}+1)\delta(t-t'), \qquad (11)$$

with $\bar{n}=\left[\exp(\hbar\omega_m/k_BT)-1\right]^{-1}$ the mean thermal phonon number, which is assumed to stay at the same environmental temperature $T$. This means that the evolution of the mechanical resonator is a Markovian process.

Once the covariance matrix $V$ is obtained, one can then calculate the entanglement between the two cavity modes and we adopt the logarithmic negativity [41], which is defined as

$$E_N = \max\left[0,-\ln 2\tilde{\nu}_-\right], \qquad (12)$$

where $\tilde{\nu}_- = \min \mathrm{eig}|i\Omega_2\tilde{V}_c|$ ($\Omega_2 = \oplus_{j=1}^2 i\sigma_y$ is the so-called symplectic matrix, $\sigma_y$ is the y-Pauli matrix and $\oplus$ denotes direct sum of matrices) is the minimum symplectic eigenvalue of the covariance matrix $\tilde{V}_c = P_{1|2}V_cP_{1|2}$, with $V_c$ the 4×4 covariance matrix related to the two cavity modes and the $P_{1|2}=\mathrm{diag}(1,1,1,-1)$ is the matrix that inverts the sign of phase of cavity mode 2, which realizes partial transposition at the level of covariance matrices [42].

## 3. Numerical results and discussions

In this section, we present the numerical results of the steady-state entanglement between two cavity modes focusing on the effects of the OPA. For such a system without OPA, the entanglement properties of two cavity modes have been investigated [12-14]. A judicious choice of detunings is vital to achieve the two-mode squeezed state of the cavity modes. Following Refs. [12-14], we set $\Delta_1 = \omega_m$ and $\Delta_2 = -\omega_m$, i.e., cavity mode 1 (2) is driven at the red (blue) sideband associated with the mechanical resonator, and we assume also that the system is in the resolved sideband limit, i.e., $\omega_m \gg \kappa \gg \gamma_m$, which requires that the cavity is of high finesse. Under these conditions, the entanglement of two cavity modes can be efficiently generated: the two-mode squeezing interaction driven by the laser at the blue sideband generates entanglement between the mechanical resonator and cavity mode 2, and the beam-splitter interaction driven by the laser at the red sideband then transfers the state of the mechanical resonator to cavity mode 1. By exchanging the roles of the cavity modes and the mechanical mode, similar mechanism can be used to prepare two-mode squeezed states of two mechanical resonators interacting with one cavity mode [26].

Figure 2 shows the steady-state entanglement between two cavity modes as a function of the phase $\theta$ of the driving field on the OPA. It shows that the optimal phase for the entanglement is $\theta = \pi/2$ for $\theta \in [0,\pi]$, while for $\theta \in [\pi, 2\pi]$, the system starts to be unstable which we will not consider. We have employed the following parameters [15, 43]: the mechanical resonator with effective mass $m = 5$ ng, frequency $\omega_m/2\pi = 10$ MHz and damping rate $\gamma_m/2\pi = 100$ Hz, the cavity with length $L = 5$ mm, finesse $F = 10^5$ (corresponding to $\kappa = 0.94$ MHz), and wavelengths about 1064 nm, two driving lasers with powers $P_1 = 100$ mW and $P_2 = 80$ mW, and detunings $\Delta_{1,2} = \pm \omega_m$. The OPA is known to generate squeezing of the optical field and the degree of squeezing is proportional to the nonlinear gain of the OPA. This is clearly shown in Figure 3, in which we plot the ratio of the variance of two quadrature fluctuations $\langle \delta X_1^2 \rangle / \langle \delta Y_1^2 \rangle$ for cavity mode 1 and $\langle \delta Y_2^2 \rangle / \langle \delta X_2^2 \rangle$ for cavity mode 2 (the variance denotes noise while the ratio not equal to 1 reflects squeezing) as a function of the nonlinear gain $G$. When $G=0$, the OPA is not working so that the ratio equal to 1 corresponding to a thermal state of the cavity field due to the interaction with the mirror which is in thermal equilibrium with the environment. We see that cavity mode 1 is phase squeezing while mode 2 is amplitude squeezing, and the two modes have the same degree of squeezing. This is due to the phase difference of two driving fields on the OPA and the assumed identical nonlinear gain $G$. This single-mode squeezing is helpful to enhance the two-mode entanglement, as shown in Figure 4. We see that the OPA can significantly enhance the entanglement and the maximum of the entanglement with OPA increases by 104% for $T=10\,\text{mK}$, 96% for $T=100\,\text{mK}$, and 55% for $T=1\,\text{K}$, compared to the value without OPA (i.e. when $G$=0). The entanglement degrades when the nonlinear gain $G$ takes large values. This can be explained in the following way: due to the different detunings $\Delta_{1,2}=\pm\omega_m$, as $G$ grows $\alpha_{s1}$ increases while $\alpha_{s2}$ decreases, leading to large difference between $G_1$ and $G_2$ when $G$ is large. Since the OPA results in a phase (amplitude) squeezed cavity mode 1 (2), this implies $\langle\delta X_1^2\rangle$ is (much) larger than $\langle\delta X_2^2\rangle$ for large $G$. Due to these two facts, when $G$ is large the optical noise becomes a significant effective thermal bath for the mechanical mode (see the second equation in Eq. (6) accounting for the mechanical momentum fluctuation), leading to the degradation of entanglement between two optical modes. Therefore, an optimal $G$ exists corresponding to maximum entanglement as a result of the balance between two effects of the OPA: entanglement enhancement at moderate values of $G$ and entanglement degradation at large values of $G$. The enhancement decreases as the temperature rises and the optimal value of $G$ for the entanglement shows similar behavior, as shown in Figure 5. This means that our scheme is preferred to work at cryogenic temperatures.

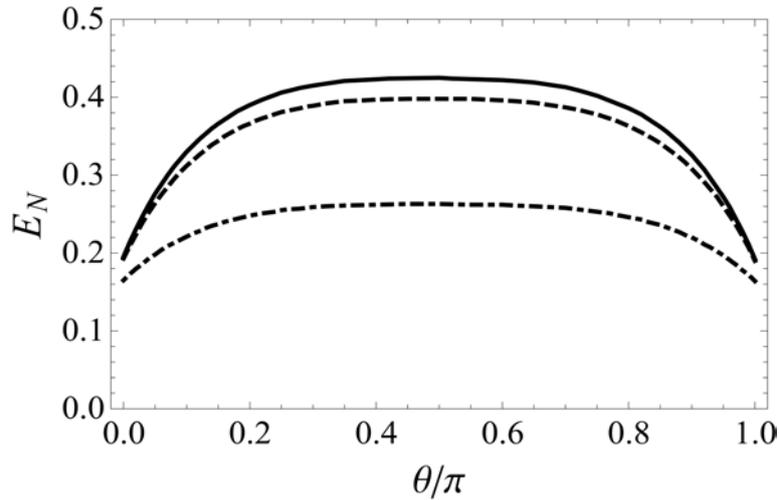

**Figure 2.** The entanglement of two cavity modes $E_N$ as a function of the phase $\theta$ for various temperatures and the corresponding optimal values of the nonlinear gain: $T = 10$ mK and $G = 5.6\,\kappa$ (solid line), $T = 100$ mK and $G = 5\,\kappa$ (dashed line), $T = 1$ K and $G = 3\,\kappa$ (dot-dashed line). See text for the other parameters.

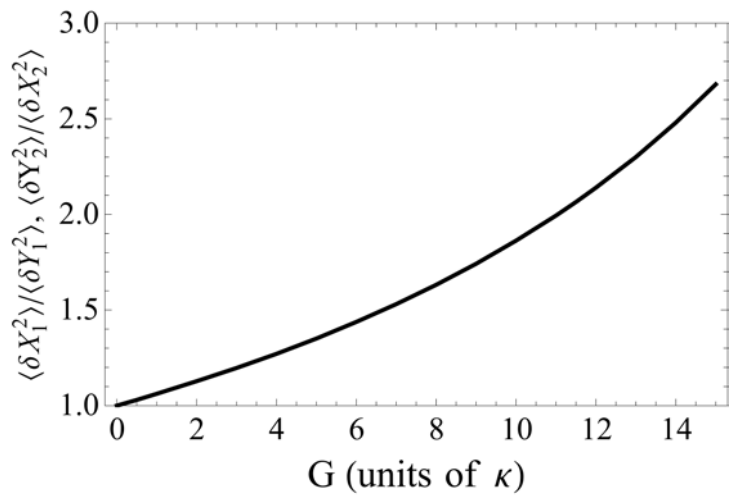

**Figure 3.** $\langle \delta X_1^2 \rangle / \langle \delta Y_1^2 \rangle$ ($\langle \delta Y_2^2 \rangle / \langle \delta X_2^2 \rangle$) of cavity mode 1 (2) versus the nonlinear gain $G$. We take $T = 10$ mK and $\theta = \pi/2$. The other parameters are given in text, which correspond to $\kappa = 0.94$ MHz.

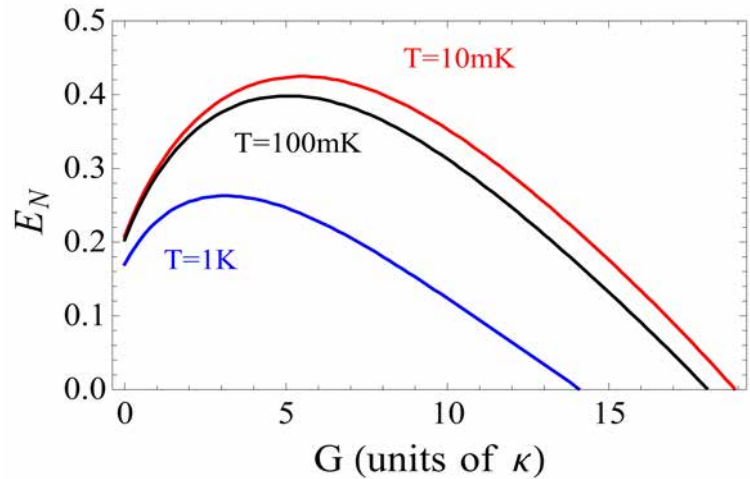

**Figure 4.** The entanglement $E_N$ versus the nonlinear gain $G$ for various system temperatures. The other parameters are as in Fig. 3.

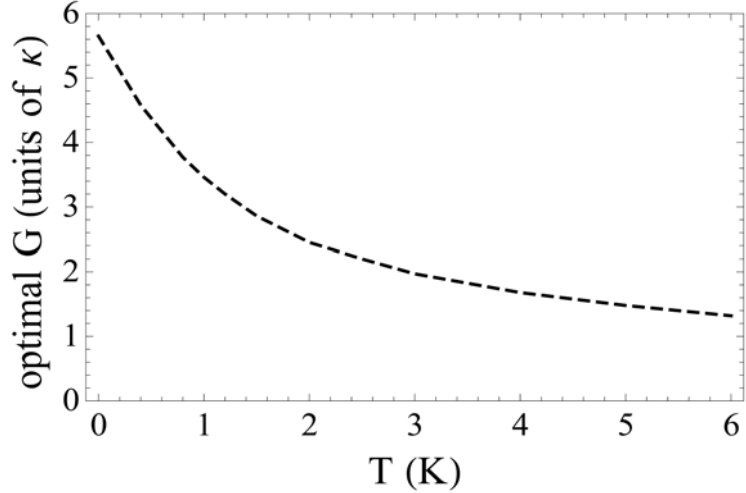

**Figure 5.** The optimal nonlinear gain $G$ for the entanglement as a function of temperature $T$. The other parameters are as in Fig. 3.

## 4. Conclusions

We have studied the effect of the OPA on the improvement of steady-state entanglement between two cavity modes which jointly interact with a mechanical resonator. The OPA generates squeezed cavity fields leading to a significant improvement of the two-mode entanglement, especially for the system at low temperatures. However, there is no simple correspondence between the single-mode squeezing and the two-mode entanglement. An optimal nonlinear gain of the OPA exists, depending on the system temperatures, which gives rise to the maximum entanglement. Therefore, the OPA with a tunable nonlinear gain, realized by adjusting the power of the driving field, is vital for optimizing the entanglement. Although the present study is focused on optical entanglement, our protocol can also be applied to enhance the entanglement between an optical mode and a microwave mode [11], where the OPA is used to squeeze only the optical field. We note that, as an alternative, one could consider using a feedback scheme [44], which can also significantly improve the entanglement between two optical modes.

# Acknowledgments

This work has been supported by the Major State Basic Research Development Program of China, the National Natural Science Foundation of China (Grant Nos: 11504218, 11634008, 11674203, 61108003, 61227902), Natural Science Foundation of Shanxi Province, China (2013021005-2) and National Key Research and Development Plan (2016YFA0301404).

# References

[1] Nielsen M A and Chuang I L 2000 Quantum Computation and Quantum Information (Cambridge: Cambridge University Press)
[2] Leggett A J 1980 Foundations of Quantum Mechanics *Prog. Theor. Phys.* **69** 80
    Leggett A J 2002 Testing the limits of quantum mechanics: motivation, state of play, prospects *J. Phys. Condens. Matter* **14** R415
[3] Zurek W H 1991 Decoherence and the Transition from Quantum to Classical *Phys. Today* **44** 36
    Zurek W H 2003 Decoherence, einselection, and the quantum origins of the classical *Rev. Mod. Phys.* **75** 715
[4] Bassi A, Lochan K, Satin S, Singh T P and Ulbricht H 2013 Models of wave-function collapse, underlying theories, and experimental tests *Rev. Mod. Phys.* **85** 471
[5] Belli S *et al* 2016 Entangling macroscopic diamonds at room temperature: Bounds on the continuous-spontaneous-localization parameters *Phys. Rev. A* **94** 012108
[6] Zhang J, Zhang T and Li J 2017 Probing Spontaneous Wave-Function Collapse with Entangled Levitating Nanospheres *Phys. Rev. A* **95** 012141
[7] Aspelmeyer M, Kippenberg T J and Marquardt F 2014 Cavity optomechanics *Rev. Mod. Phys.* **86** 1391
[8] Hammerer K, Genes C, Vitali D, Tombesi P, Milburn G, Simon C and Bouwmeester D 2012 Nonclassical States of Light and Mechanics arXiv: 1211.2594
[9] Bose S, Jacobs K and Knight P L 1997 Preparation of nonclassical states in cavities with a moving mirror *Phys. Rev. A* **56** 4175
[10] Paternostro M, Vitali D, Gigan S, Kim M S, Brukner C, Eisert J and Aspelmeyer M 2007 Creating and Probing Multipartite Macroscopic Entanglement with Light *Phys. Rev. Lett.* **99** 250401
[11] Barzanjeh S, Vitali D, Tombesi P and Milburn G J 2011 Entangling optical and microwave cavity modes by means of nanomechanical resonator *Phys. Rev. A* **84** 042342
[12] Tian L 2013 Robust Photon Entanglement via Quantum Interference in Optomechanical Interfaces *Phys. Rev. Lett.* **110** 233602
[13] Wang Y D and Clerk A A 2013 Reservoir-Engineered Entanglement in Optomechanical Systems *Phys. Rev. Lett.* **110** 253601
[14] Kuzyk M C, van Enk S J and Wang H 2013 Generating robust optical entanglement in

    weak-coupling optomechanical systems *Phys. Rev. A* **88** 062341
[15] Vitali D *et al* 2007 Optomechanical Entanglement between a Movable Mirror and a Cavity Field *Phys. Rev. Lett*. **98** 030405
[16] Mari A and Eisert J 2009 Gently Modulating Optomechanical Systems *Phys. Rev. Lett*. **103** 213603
[17] Hofer S G, Wieczorek W, Aspelmeyer M and Hammerer K 2011 Quantum entanglement and teleportation in pulsed cavity optomechanics *Phys. Rev. A* **84** 052327
[18] Mancini S, Giovannetti V, Vitali D and Tombesi P 2002 Entangling Macroscopic Oscillators Exploiting Radiation Pressure *Phys. Rev. Lett*. **88** 120401
[19] Hartmann M J and Plenio M B 2008 Steady State Entanglement in the Mechanical Vibrations of Two Dielectric Membranes *Phys. Rev. Lett*. **101** 200503
[20] Zhang J, Peng K and Braunstein S L 2003 Quantum-state transfer from light to macroscopic oscillators *Phys. Rev. A* **68** 013808
[21] Pirandola S, Vitali D, Tombesi P and Lloyd S 2006 Macroscopic Entanglement by Entanglement Swapping *Phys. Rev. Lett*. **97** 150403
[22] Borkje K, Nunnenkamp A and Girvin S M 2011 Proposal for Entangling Remote Micromechanical Oscillators via Optical Measurements *Phys. Rev. Lett*. **107** 123601
[23] Tan H, Li G and Meystre P 2013 Dissipation-driven two-mode mechanical squeezed states in optomechanical systems *Phys. Rev. A* **87** 033829
[24] Schmidt M, Ludwig M and Marquardt F 2012 Optomechanical circuits for nanomechanical continuous variable quantum state processing *New J. Phys*. **14** 125005
[25] Woolley M J and Clerk A A 2014 Two-mode squeezed states in cavity optomechanics via engineering of a single reservoir *Phys. Rev. A* **89** 063805
[26] Li J, Moaddel Haghighi I, Malossi N, Zippilli S and Vitali D 2015 Generation and detection of large and robust entanglement between two different mechanical resonators in cavity optomechanics *New J. Phys*. **17** 103037
    Li J, Li G, Zippilli S, Vitali D and Zhang T 2016 Enhanced entanglement of two different mechanical resonators via coherent feedback arXiv: 1610.07261
[27] Genes C, Vitali D and Tombesi P 2008 Emergence of atom-light-mirror entanglement inside an optical cavity *Phys. Rev. A* **77** 050307 (R)
[28] Rogers B, Paternostro M, Palma G M and De Chiara G 2012 Entanglement control in hybrid optomechanical systems *Phys. Rev. A* **86** 042323
[29] He Q and Ficek Z 2014 Einstein-Podolsky-Rosen paradox and quantum steering in a three-mode optomechanical system *Phys. Rev. A* **89** 022332
[30] Zhang J, Zhang T, Xuereb A, Vitali D and Li J 2015 More nonlocality with less entanglement in a tripartite atom-optomechanical system *Ann. Phys*. **527** 147
[31] Huang S and Agarwal G S 2009 Enhancement of cavity cooling of a micromechanical mirror using parametric interactions *Phys. Rev. A* **79** 013821
[32] Huang S and Agarwal G S 2009 Normal-mode splitting in a coupled system of a nanomechanical oscillator and a parametric amplifier cavity *Phys. Rev. A* **80** 033807
[33] Lü X Y, Wu Y, Johansson J R, Jing H, Zhang J and Nori F 2015 Squeezed Optomechanics with Phase-Matched Amplification and Dissipation *Phys. Rev. Lett*. **114** 093602
[34] Peano V, Schwefel H G L, Marquardt C and Marquardt F 2015 Intracavity Squeezing Can Enhance Quantum-Limited Optomechanical Position Detection through Deamplification

    *Phys. Rev. Lett.* **115** 243603
[35] Agarwal G S and Huang S 2016 Strong mechanical squeezing and its detection *Phys. Rev. A* **93** 043844
[36] Mi X, Bai J and Song K H 2013 Robust entanglement between a movable mirror and a cavity field system with an optical parametric amplifier *Eur. Phys. J. D* **67** 115
[37] Hu C S, Huang X R, Shen L T, Yang Z B and Wu H Z 2016 Enhancement of entanglement in distant micromechanical mirrors using parametric interactions arXiv:1606.08946
[38] Xuereb A, Barbieri M and Paternostro M 2012 Multipartite optomechanical entanglement from competing nonlinearities *Phys. Rev. A* **86** 013809
[39] Giovannetti V and Vitali D 2001 Phase-noise measurement in a cavity with a movable mirror undergoing quantum Brownian motion *Phys. Rev. A* **63** 023812
[40] Gröblacher S, Hammerer K, Vanner M R and Aspelmeyer M 2009 Observation of strong coupling between a micromechanical resonator and an optical cavity field *Nature* **460** 724
[41] Eisert J 2001 PhD Thesis, University of Potsdam
    Vidal G and Werner R F 2002 Computable measure of entanglement *Phys. Rev. A* **65** 032314
    Plenio M B 2005 Logarithmic Negativity: A Full Entanglement Monotone That is not Convex *Phys. Rev. Lett.* **95** 090503
[42] Simon R 2000 Peres-Horodecki Separability Criterion for Continuous Variable Systems *Phys. Rev. Lett.* **84** 2726
[43] Gigan S *et al* 2006 Self-cooling of a micromirror by radiation pressure *Nature* **444** 67
[44] Asjad M, Tombesi P and Vitali D 2016 Feedback control of two-mode output entanglement and steering in cavity optomechanics *Phys. Rev. A* **94** 052312